\documentclass[preprint,showpacs,preprintnumbers,nofootinbib]{revtex4}
\usepackage{graphicx}
\usepackage{epstopdf}
\usepackage{dcolumn}
\usepackage{bm}
\usepackage{axodraw}

\newcommand{\nn}{\nonumber}

\newcommand{\be}{\begin{eqnarray}}
\newcommand{\ee}{\end{eqnarray}}
\catcode`\@=11
\def\lsim{\mathrel{\mathpalette\@versim<}}
\def\gsim{\mathrel{\mathpalette\@versim>}}
\def\@versim#1#2{\vcenter{\offinterlineskip
\ialign{$\m@th#1\hfil##\hfil$\crcr#2\crcr\sim\crcr } }}
\catcode`\@=12

\begin{document}
\def\thefootnote{\fnsymbol{footnote}}
\begin{flushright}
KANAZAWA-10-05  \\
 DO-TH 10/11\\
July, 2010
\end{flushright}
\vspace*{0.5cm}
\begin{center}
{\LARGE\bf 
Large loop effects of extra  SUSY Higgs 
\\

\vspace{0.2cm}

 doublets  to CP violation
in  $B^0$ mixing
}\\
\vspace{1. cm}

{\Large Jisuke Kubo}
\footnote[1]{e-mail:~jik@hep.s.kanazawa-u.ac.jp}
\vspace {0.5cm}\\
{\it Institute for Theoretical Physics, Kanazawa University,\\
        Kanazawa 920-1192, Japan}
\vspace {0.5cm}\\
and 
\vspace {0.5cm}\\
{\Large Alexander Lenz}
\footnote[2]{e-mail:~alexander.lenz@physik.uni-regensburg.de}
\vspace {0.5cm}\\
{\it Institut f\"ur Physik, Technische Universit\"at Dortmund, \\  
        D--44221 Dortmund, Germany}
\\  
{\it Institut f{\"u}r Theoretische Physik, Universit{\"a}t Regensburg,\\
        D-93040 Regensburg, Germany}\\
\end{center}
\vspace{1cm}
{\Large\bf Abstract}\\
We consider more than one pair of $SU(2)_L$ doublet Higgs supermultiplets
in a generic supersymmetric extension of the standard model, and
calculate their one-loop contributions to the soft mass insertions 
$\delta_{LL}$ etc. We find that if  large supersymmetry breaking
in this sector is realized, the loop effects can give rise to 
large contributions to the soft mass insertions, meaning that they 
can generate large FCNCs and CP violations.
\\
We apply our result to a recently proposed model based on the 
discrete $Q_6$ family group, and calculate the non-diagonal matrix 
element $M_{12}$  of the neutral meson systems.
We focus our attention on  the extra phases $\phi_{d,s}^\Delta$ in $B_{d,s}$-mixing
and flavor-specific CP-asymmetries $a_{sl}^{d,s}$ in neutral $B$ decays
and obtain values that can be about one order of magnitude larger than the
standard model predictions. Our final results are comparable with the recent 
experimental observations at D0 and CDF, but they are still about a factor of
5 smaller than the recently measured dimuon asymmetry from D0.

\newpage
\setcounter{footnote}{0}
\def\thefootnote{\arabic{footnote}}
\section{Introduction}

The CKM mechanism \cite{CKM} has been tested and confirmed to a high accuracy  
as the dominant source of flavor violation and CP violation in the standard model 
(SM), see e.g. \cite{CKMfitter,UTfit}. Despite this success it is well-known 
that the amount of CP violation present in the SM is not sufficient 
\cite{notenoughCP} to explain the baryon asymmetry in the universe \cite{Sakharov:1967dj}.
Moreover in the last years also some hints for
deviations of the CKM picture were accumulated.
In particular the recent measurement of the CP-violating dimuon asymmetry by the D0 collaboration
\cite{Abazov:2010hv} gained a lot of attention.
The measured value
\begin{equation}
A_{sl}^b = -(9.57 \pm 2.51 \pm 1.46) \cdot 10^{-3} \, ,
\label{d0}
\end{equation}   
is a factor of 42 larger than the SM prediction
\cite{LN2006}
\begin{equation}
A_{sl}^b = -(2.3^{+0.5}_{-0.6}) \cdot 10^{-4} \, .
\label{asl-sm}
\end{equation}   
The statistical significance of this deviation is 3.2 $\sigma$. 
Since this large deviation might hint to a sizeable new source of CP violation, 
needed to solve the problem of the baryon asymmetry, the D0 measurement
resulted already in many theoretical papers investigating different new physics models
\cite{NPdimuon}, for earlier works on large CP violation in $B_s$ mixing see e.g.
\cite{NPphisold,Kawashima:2009jv,Babu:2009nn}.

The SM and its minimal supersymmetric extension
(MSSM) with softly broken supersymmetry have the minimal structure of the Higgs sector.
In the SM this minimality is the main reason  that the flavor-changing-neutral current (FCNC)
as well as the CP-violating processes are highly suppressed.  
However, the minimality of the Higgs sector in the MSSM does not help suppressing
FCNC and CP violation at all, and this gives arise to
the well-known  SUSY flavor and CP problem
\cite{Hall:1985dx,Gabbiani:1988rb,Ellis:1981ts,Ellis:1982tk,Bertolini:1990if,Branco:1994eb,Gabbiani:1996hi,Misiak:1997ei,Abel:2001vy}.
Therefore, one is led to consider another
mechanism to suppress FCNC and CP violation in supersymmetric 
extensions of the SM.  
A natural assumption is that spontaneous CP violation  or its modification 
is responsible for the small CP violation in the MSSM.
However, spontaneous CP violation in the MSSM
does not occur, unless one extends the Higgs sector
to a non-minimal form \cite{Romao:1986jy}. Moreover, just adding 
more $SU(2)_L$ doublet Higgs
supermultiplets does not help; one should introduce 
a certain set of SM singlet Higgs bosons \cite{Romao:1986jy}.
However, an extension to introduce more than one pair of $SU(2)$ doublet Higgs supermultiplets 
might suffer from two major problems: 
\begin{itemize}
\item[(i)]   it can destroy the successful gauge coupling unification, and 
\item[(ii)]  large  tree-level FCNCs  can be  present.
\end{itemize}
We will ignore problem (i) in this work, while we will solve problem (ii) 
by introducing a flavor symmetry;  a symmetry-based mechanism 
to suppress FCNCs was considered in the literature e.g. in 
 \cite{Dine:1993np}-\cite{Kifune:2007fj}, 
 \cite{ Kawashima:2009jv,Babu:2009nn}\footnote{ Recently,
a considerable attention has been given to the idea of incorporating a non-abelian
flavor symmetry into a GUT\cite{GUT}.}.
Note, however,  that if 
CP violation in $B^0$ mixing should turn out to be  large as the D0 measurement 
(\ref{d0}) is suggesting,
we are caught in a  dilemma between suppressed and large CP violation.
In this paper we would like to address this dilemma (see also S.~King
of \cite{NPdimuon}, who has also addressed the problem in a similar framework very recently).

First we consider  more than one pair of $SU(2)_L$ doublet Higgs supermultiplets
in a generic supersymmetric extension of the SM.
We then calculate the one-loop contributions of the extra heavy Higgs multiplets
to the soft mass insertions $\delta$'s \cite{Hall:1985dx,Gabbiani:1988rb}.
We find  that the loop effects can give rise to large contributions to the soft 
mass insertions. That is, the loop effects can generate large FCNCs and CP violations
in such models.
We then apply our general result to a specific model, in which the problem (ii) is
overcome by the  flavor symmetry $Q_6$ \cite{Babu:2004tn}, and 
investigate the one-loop effects mentioned above   on the dimuon asymmetry 
and related observables. 
The Higgs sector of this model consists of six $SU(2)_L$ doublets, 
three for the up-quark sector and
three for the down-quark sector; the three doublets of each sector form a three-dimensional reducible
representation  ${\bf 1}+{\bf 2}$  of $Q_6$.
In the end four of the six $SU(2)_L$ doublets are super-heavy $\gsim$ few TeV
(which comes  from the FCNC constraints in the mixing of the neutral meson systems
\cite{Kifune:2007fj,Babu:2009nn}),
and two of them form the pair of the MSSM Higgs supermutiplets.
Tree-level contributions to the semileptonic asymmetries,
due to  the exchange of the extra heavy neutral Higgs bosons,
were discussed in \cite{Kawashima:2009jv}. There it was found that the small standard model
expectations for semileptonic CP asymmetries can be enhanced by up to one order
of magnitude.
In this paper we assume that the extra heavy
Higgs bosons are so heavy that the tree-level contributions can be neglected,
and that the extra FCNCs and CP violations come only from the SUSY breaking sector.
We determine under this assumption in this model the non-diagonal matrix element $M_{12}$  
of the neutral meson systems and investigate the possible size of
a new CP-violating phase as well as the possible size of semileptonic CP-asymmetries.
Our results are finally compared with the recent  measurements, in particular
the dimuon asymmetry of D0 \cite{Abazov:2010hv}.

\section{New Physics in $B^0$ mixing}

The mixing of neutral mesons is governed by the famous box diagrams. The dispersive part,
denoted by $M_{12}$ is expected to be very sensitive to new physics, while for the
absorptive part, denoted by $\Gamma_{12}$, new contributions are expected to be below
the hadronic uncertainties \footnote{This statement was recently questioned in the literature.
A more detailed discussion of it, will be given in \cite{LR2010}.}. 
Therefore one can write generally \cite{LN2006} 
in the presence of new physics
\begin{eqnarray}
M_{12}^q & = & M_{12}^{SM,q} \cdot \Delta_q \, , \hspace{1cm}  \Delta_q = | \Delta_q| e^{i \phi_q^\Delta} \, ,
\label{M12allg}
\\
\Gamma_{12}^q & = & \Gamma_{12}^{SM,q} \, .
\label{G12allg}
\end{eqnarray}
$q=s$ denotes the $B_s$-system and $q=d$ denotes the $B_d$-system.
Defining the phase $\phi_q$ as
\begin{equation}
\phi_q = \arg \left( - \frac{M_{12}^q}{\Gamma_{12}^q} \right) \, ,
\end{equation}
we get \cite{LN2006}  the following general expression for the flavor-specific CP-asymmetries (sometimes also
called semileptonic CP-asymmetries)
in the presence of new physics
\begin{eqnarray}
a_{sl}^q & = & \mbox{Im} \left( \frac{\Gamma_{12}^q}{M_{12}^q} \right)
= \frac{|\Gamma_{12}^{q}|}{|M_{12}^{SM,q}|} \cdot \frac{\sin (\phi_q^{SM} + \phi_q^\Delta)}{|\Delta_q|} \, .
\label{NPinasl}
\end{eqnarray} 
The above formula holds independent of the assumption, that there is almost no new physics
possible in $\Gamma_{12}$.
In the SM the CP-violating phases $\phi_q$ and the semileptonic CP-asymmetries are small, 
one gets   \cite{LN2006} (based on the results of \cite{BBD06,BBGLN98,BBGLN02,BBLN03,rome03})
\begin{eqnarray}
a_{sl}^d = (-4.8^{+1.0}_{-1.2}) \cdot 10^{-4} \, , 
& \hspace{1cm}& 
\phi_d^{SM} = -0.091^{+0.026}_{-0.038} = {-5.2^\circ}^{+1.5^\circ}_{-2.1^\circ}\, ,
\label{phid}
\\
a_{sl}^s = (2.06\pm 0.57) \cdot 10^{-5} \, ,
&& 
\phi_s^{SM} = (4.2 \pm 1.4) \cdot 10^{-3} = {0.24^\circ} \pm 0.08^\circ \, .
\label{phis}
\end{eqnarray}
D0 measured \cite{Abazov:2010hv} a linear combination of the semileptonic CP-asymmetries in the $B_d$ and 
in the $B_s$ system
\begin{equation}
A_{sl}^b = (0.494\pm0.043) \cdot a_{sl}^s + (0.506\pm0.043) \cdot a_{sl}^d \, .
\label{Asl}
\end{equation}
The experimental central value turned out to be a factor of 42 larger than the SM
 expectation for $A_{sl}^b$.
Using the experimental value for $a_{sl}^d = -0.0047 \pm 0.0046$ from \cite{HFAG} one derives \cite{Abazov:2010hv}
a bound on $a_{sl}^s$:
\begin{equation}
a_{sl}^s = ( - 14.6 \pm 7.5) \cdot 10^{-3} \, .
\label{aslsexp}
\end{equation}
Inserting this value in Eq.(\ref{NPinasl}) we get with the results from  \cite{LN2006}
\begin{equation}
\sin (\phi_s^{SM} + \phi_s^\Delta) = -(2.9 \pm 1.5)  \cdot |\Delta_s|.
\end{equation}
Assuming however, that there is no new physics in $B_d$-mixing one gets instead
\begin{equation}
a_{sl}^s = (-19 \pm 10) \cdot 10^{-3} \, \, 
\Rightarrow 
\, \, \sin (\phi_s^{SM} + \phi_s^\Delta) = -(3.8 \pm 2.0) \cdot |\Delta_s|.
\end{equation}
Using the fact that $|\Delta |$ is closed to one (see e.g. \cite{LN2006})
we get in both cases unphysical values for $ \sin (\phi_s^{SM} + \phi_s^\Delta) $.
This problem will be discussed in detail in \cite{LR2010}, here we simply
assume that the current data hint for a large value of  $\phi_s^\Delta$
compared to the SM angle $\phi_s^{SM}$. This also holds if we combine the D0 dimuon asymmetry
with previous direct determinations of semileptonic CP-asymmetries \cite{asldirect}. 
\\
If a non-vanishing value of $\phi_s^\Delta$ is realized in nature, this would also be
visible in the angular analysis of the decay $B_s \to J / \Psi \Phi$ \cite{FDN}. 
In the SM one extracts in this decay the angle $- 2 \beta_s \approx -2.2^\circ$ (for the notation 
see e.g. Noted added in \cite{noteadded}).
If new physics is only present in the $B_s$-mixing and not in the $B_s \to J / \Psi \Phi$ decay one extracts 
instead the angle $-2 \beta_s + \phi_s^\Delta$.
Current data \cite{Aaltonen:2007he,2008fj,combi} for $B_s \to J / \Psi \Phi$ also hint to a non-vanishing value of  
$ \phi_s^\Delta$\footnote{A new result from CDF \cite{CDF} was presented at FPCP 2010 
giving a 1-sigma range of $\phi_s^\Delta \in [0,-1]$, being perfectly consistent with  the SM
($\phi_s^\Delta = 0$), but also 
with a large deviation from the SM($\phi_s^\Delta = -1 \approx -57^\circ$).}, 
which points to the same direction as the value of the 
semileptonic CP-asymmetries measured by D0 \cite{Abazov:2010hv}.
Possible problems using this extraction for the CP-violating phase in $B_s$ mixing are 
discussed in detail in \cite{LR2010}.

\section{General formula}

Consider the superpotential
\be
W
&=&
\hat{Y}_{ij}^{uI} Q_{i} U_{j}^c  H^u_I+
\hat{Y}_{ij}^{dI} Q_{i} D_{j}^c  H^d_I+
\mu^{IJ} H^u_I H^d_J~ .
\label{superP}
\ee
Here $SU(2)_L$ doublets
of the quark and Higgs supermultiplets
are denoted by $Q , H^u$ and  $ H^d$, 
respectively. The indices $I$ and $J$ indicate different kinds of the Higgs $SU(2)_L$ doublets.
Similarly, $U^c$ and  $D^c$ stand for $SU(2)_L$ singlets 
of the quark supermultiplets.
We denote the component fields by a small letter along with
a $\sim$ for the scalar quarks and higgsinos, and $(\hat{Y})^*=Y$.
($Y$ is the Yukawa coupling in our notation, i.e. $Y \bar{q}_L q_R$.)
The $SU(2)$ components of the Higgs fields are
\be
h_I^u &=&(  \phi_I^{u+}~,~\phi_I^{u0})~,~
h_I^d=(  \phi_I^{d0}~,~\phi_I^{d-})~.
\label{su2-comp}
\ee
To compute the corrections to the soft mass 
insertions $(\delta_{ij})_{LL,\rm etc}$,
which will be defined in (\ref{Delta1}),
we have to compute the corrections to 
the  squark masses. In the following  calculations we consider only
the insertions $(\delta_{12}^d)_{LL,LR}$ for the down-type scalar quarks, and 
moreover neglect the $Y^d$'s except for the tree-level $\delta_{LR}$.
Quark and squark masses are also neglected.
Calculation of other types of the insertions can be done in a similar way.
There will be (i) tree-level
contributions to $\delta_{LR}$ coming from the fact that there are 
more than one pair of Higgs doublets.
Then (ii)   diagrams of Fig.~1 with heavy Higgs fields in the loop,
and two types of loop diagrams; (iii) those with the heavy Higgs bosons and 
squarks in the loop  as shown in Fig.~2 (a) and (vi) 
those with the higgsinos and quarks in the loop  as shown in  Fig.~2 (b).
We obtain the following results:

\vspace{0.3cm}
\noindent
(i) Tree-level contribution to $(\delta_{ij}^{d})_{LR}$:\\
The relevant Lagrangian is
\be
{\cal L}_{\mu} &=&-\left(\phi_I^{u0} \mu^{IJ} Y_{ij}^{dJ}\right)\tilde{d}_{Li}^*
\tilde{d}_{Rj}+h.c.~,
\ee
which yields
\be
(\delta_{ij}^d)_{LR}(\mu) &=& \left(\frac{<\phi_I^{u0}> 
\mu^{IJ}}{m_{\tilde{d}}^2} \right)
\left[(U_L^d)^\dag Y^{dJ } U_R^d)\right]_{ij}~,
\label{dLR-mu}
\ee
where $m_{\tilde{d}}$  is the average squark mass, and
$U'$s are  unitary matrices that diagonalize the   down-type
quark mass matrix, i.e.
\be
(U_L^{d})^\dag {\bf m}_{d }U_R^{d}&=& 
\mbox{diag}.(m_{d},m_{s},m_{b})~.
\label{uRuL}
\ee
There are besides the usual contribution to
$\delta_{LR}$ coming from the $A$ terms, which we have not
included. The contribution (\ref{dLR-mu}) exists only if there are more than one pair of
Higgs doublets.

\vspace{0.3cm}
\noindent
(ii)  Quartic coupling contribution  to $(\delta_{ij}^d)_{LL}$ (Fig.~1):\\
The quartic couplings are given by
\be
{\cal L}_{\mbox{quart}}&=&
-Y^{uI}_{ik} (Y^{uJ})_{kj}^\dag (\phi_I^{u+})^* 
\phi_J^{u+}~ \tilde{d}_{Li}^* \tilde{d}_{Lj} ~.
\ee
We find
\be
(\delta_{ij}^d)_{LL}(\mbox{quart})
&=&\frac{[(U_L^d)^\dag Y^{uI} (Y^{uJ})^\dag 
U_L^d ]_{ij}}{16 \pi^2 m_{\tilde{d}}^2}
 (U_c ~{\bf M}^{2,dia}_{c} ~U_c^\dag)_{JI}~,
 \label{dLL-quart}
\ee
where
\be
 \phi_I^{u+} &=&U_{c,IJ} \phi_J^{u+,dia}~,~
({\bf M}^{2,dia}_{c})_{IJ}=
(m_{c,J}^2\ln m_{c,J}^2/Q^2)\delta_{IJ}~,~
U_c^\dag~  {\bf M}^{2}_{c} ~
U_c = m_{c,J}^2\delta_{IJ}~.
\label{Mc}
\ee
${\bf M}^{2}_{c}$ is the mass matrix for the charged Higgs bosons, and
$Q$ is the renormalization scale. We have suppressed the constant terms
which, however, can be absorbed into $Q$.
We will do so for other diagrams.

\vspace{0.3cm}
\noindent
(iii) Cubic scalar couplings  to $(\delta_{ij}^d)_{LL}$
 (Fig.~2 (a)):
\be
{\cal L}_{\mbox{cs}}=Y_{ij}^{uI} \mu^{IJ} \phi_J^{d-} 
~ \tilde{d}_{Li}^* \tilde{u}_{Rj} +h.c,
\ee
which gives
\be
(\delta_{ij}^d)_{LL}(\mbox{cs})
&=&\frac{[(U_L^d)^\dag Y^{uI} (Y^{uJ})^\dag 
U_L^d]_{ij}}{16 \pi^2 m_{\tilde{d}}^2}
(\mu U_c^* ~{\bf L}^{2,dia}_{c} ~U_c^T \mu^\dag)_{JI}~,
\label{dLL-cs}
\ee
where 
$({\bf L}^{2,dia}_{c})_{IJ}=(\ln m_{c,J}^2/Q^2)\delta_{IJ}$, and
$U_c$  and $m_{c,J}$ are  defined in (\ref{Mc}).
Here we have neglected the mass of $\tilde{u}_R$.

\vspace{0.3cm}
\noindent
(vi) Higgsino loop to $(\delta_{ij}^d)_{LL}$ (Fig.~2 (b)):
\be
{\cal L}_{h}=-Y_{ij}^{uI} 
~ \tilde{d}_{Li}^* \overline{\tilde{h}}_I^{u+}u_{Rj} +h.c.
\ee
The expression for $(\delta_{ij}^d)_{LL}(h)$ is similar to
the quartic  coupling contribution (\ref{dLL-quart}). We find,  neglecting the quark masses,
\be
(\delta_{ij}^d)_{LL}(h)
&=&-2\frac{[(U_L^d)^\dag Y^{uI} (Y^{uJ})^\dag 
U_L^d]_{ij}}{16 \pi^2 m_{\tilde{d}}^2}
 \left[U_h ~{\bf M}^{2F,dia}_{h}~U_h^\dag\right]_{JI}~,
 \label{dLL-f}
\ee
where $\tilde{h}_I^{u+} =U_{h,IJ} \tilde{h}_J^{u+,dia}$, and
\be
({\bf M}^{2F,dia}_{h})_{IJ}& =&(m_{h,J}^2\ln m_{h,J}^2/Q^2)\delta_{IJ}~,~
(U_h^\dag~  {\bf M}_{h}^F  U_h)_{IJ} = m_{h,J}\delta_{IJ}~.
\ee
${\bf M}_{h}^F$ is the mass matrix for the charged higgsinos.

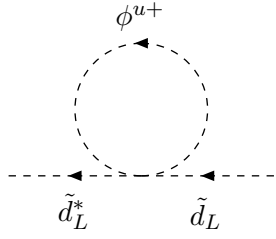
\begin{figure}[htb]
\begin{center}\begin{picture}(300,110)(0,45)
\DashArrowLine(150,50)(100,50){3}
\DashArrowLine(200,50)(150,50){3}
\Text(125,30)[b]{$\tilde{d}_L^*$}
\Text(175,30)[b]{$\tilde{d}_L$}
\DashArrowArc(150,75)(25,-90,270){3}
\Text(150,105)[b]{$\phi^{u+}$}
\end{picture}
\label{quart1}\end{center}
\caption[]{The quartic coupling contribution to $(\delta_{ij}^d)_{LL}$.}
\end{figure}

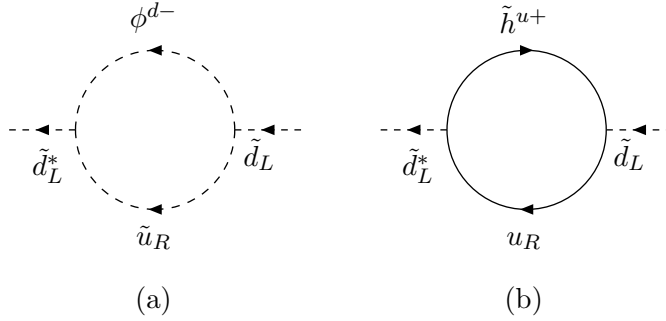
\begin{figure}[htb]
\begin{picture}(400,150)(0,-20)
\DashArrowLine(205,50)(180,50){3}
\DashArrowLine(120,50)(95,50){3}
\Text(110,30)[b]{$\tilde{d}_L^*$}
\Text(190,35)[b]{$\tilde{d}_L$}
\Text(150,5)[b]{$\tilde{u}_R$}
\Text(150,-20)[b]{(a)}
\Text(150,87)[b]{$\phi^{d-}$}
\DashArrowArc(150,50)(30,0,180){3}
\DashArrowArcn(150,50)(30,0,180){3}
\DashArrowLine(345,50)(320,50){3}
\DashArrowLine(260,50)(235,50){3}
\Text(250,30)[b]{$\tilde{d}_L^*$}
\Text(330,35)[b]{$\tilde{d}_L$}
\Text(290,5)[b]{$u_R$}
\Text(290,-20)[b]{(b)}
\Text(290,87)[b]{$\tilde{h}^{u+}$}
\ArrowArcn(290,50)(30,180,0)
\ArrowArcn(290,50)(30,0,180)
\end{picture}
\label{loop}
\caption[]{The heavy Higgs  (a) and higgsino (b) contributions to
$(\delta_{ij}^d)_{LL}$.}
\end{figure}

Before we apply the results above we make few remarks.
The infinite renormalization of the soft scalar masses do not depend
on the $\mu$  and $A$ terms to all orders in perturbation theory  
\cite{Jack:1997pa}.
Therefore, the $\mu$ parameter dependence of the infinite part 
(and hence of $\ln Q$) in $\delta$'s should be cancelled.
 However, the cancellation
of the finite part is not exact.
As we see from (\ref{dLL-cs}) - (\ref{dLL-f}), the insertions $(\delta)_{LL}$'s
explicitly depend on $\mu$ parameters
 (${\bf M}^2_c$ and   ${\bf M}^F_h$  also contain $\mu$ parameters).
Keeping this in mind,
we consider
\be
D &=&\mu_1^2 \ln m_1^2/Q^2+\mu_2^2 \ln m_2^2/Q^2
-(\mu_1^2+\mu_2^2) \ln m_3^2/Q^2
\ee
in which the renormalization scale $Q$ dependence exactly cancels. 
If all $m_i$'s are of the same 
size, $D$ is small compared to the $\mu^2$'s. 
However, if there is a large SUSY breaking
so that the mass of 
a fermionic component (higgsino) differs from that of the bosonic 
component (Higgs) by a large amount,
$D$ may become large. 
Moreover, there are terms in $\delta$'s which, instead of $\mu^2$,  are
proportional to the square of the soft scalar
masses of the Higgs bosons, which we denote generically by $m_s$.
If $m_s^2 >> \mu^2$, these $m_s^2$ terms dominate.
Then, there will be corrections
$\sim y^2(m_s^2/16 \pi^2 m_{\tilde{d}}^2) \ln m_s^2/Q^2$ to
the $\delta$'s, where $y$ stands for a generic
Yukawa coupling.  
So, if  $m_s/m_{\tilde{d}} >>1$ is realized by one reason
or another, the loop effects of the heavy Higgs
bosons to the soft mass insertions $\delta$'s may become  large.
One reason may be the following.
Flavor-changing neutral  Higgs bosons should be made heavy to
suppress tree-level FCNC processes, while  higgsinos
 should be light  because we need small $\mu$'s to suppress EDMs
which are caused by the tree-level $(\delta)_{LR}$ given in (\ref{dLR-mu}),
as was pointed out in \cite{Babu:2009nn}.
To make the Higgs bosons heavy, we have to make the corresponding soft scalar masses
large, but not the $\mu$ parameters from the reason above.
This leads to a large SUSY breaking.

\section{An application}
In this section  we would like to apply our results
of the last section to a specific model.
The model \cite{Babu:2004tn} is based on a discrete flavor symmetry
along with spontaneous CP violation.
In this model the tree-level CP violation is  suppressed 
\cite{Kawashima:2009jv,Babu:2009nn},
and it is difficult to explain  possible large CP violations, for which recently some evidence
was observed at D0 \cite{Abazov:2010hv}. We would like to see whether
the heavy Higgs fields present in the model
can help obtaining large CP violation in the
$B^0-\bar{B}^0$ mixing.
\subsection{The  model}
The model is briefly described below (the details of the model
can be found in \cite{Kajiyama:2005rk,Kifune:2007fj,Babu:2009nn}).
The $Q_6$ assignment is shown in Table \ref{assignment}.
 Here we restrict ourselves to
 the quark sector. 
\begin{table}
\begin{center}
\begin{tabular}{|c|c|c|c|c|c|c|c|c|c|c|c|}
\hline
 & $Q$ 
 & $Q_3$  
& $U^c,D^c$  
& $U^c_3,D^c_3$ 
 & $L$ & $L_3$ 
 &$E^c,N^c$ & $E_3^c$  &   $N_3^c$ 
  & $H^u,H^d$
 & $H^u_3,H^d_3$ 
\\ \hline
$Q_6$ &${\bf 2}_1$ & ${\bf 1}_{+,2}$ &
 ${\bf 2}_{2}$ &${\bf 1}_{-,1}$ &${\bf 2}_{2}$ &
${\bf 1}_{+,0}$  & ${\bf 2}_{2}$ & 
${\bf 1}_{+,0}$ & ${\bf 1}_{-,3}$ &
${\bf 2}_{2}$ & ${\bf 1}_{-,1}$  \\ \hline
\end{tabular}
\caption{ \footnotesize{The $Q_{6}$ assignment 
of the chiral matter supermultiplets, where  the group theory notation is given in 
 Ref.~\cite{Babu:2004tn}. For completeness we include
 leptons, $L, E^c$ and $N^c$. $R$ parity is also imposed.
}}
\label{assignment}
\end{center}
\end{table}
The most general $Q_6$ invariant, renormalizable
 superpotential for  the Yukawa interactions in
the quark sector yields the following Yukawa matrices \cite{Babu:2004tn}:
\be
{\bf Y}^{u1(d1)} &=&\left(\begin{array}{ccc}
0 & 0 & 0 \\
0 & 0 & Y_b^{u(d)} \\
0&  Y_{b'}^{u(d)}  & 0 \\
\end{array}\right),~
{\bf Y}^{u2(d2)} =\left(\begin{array}{ccc}
0 & 0 & Y_b^{u(d)}\\
0 & 0 & 0 \\
  -Y_{b'}^{u(d)} &0 & 0 \\
\end{array}\right),\nn\\
{\bf Y}^{u3(d3)}&=&\left(\begin{array}{ccc}
0 & Y_c^{u(d)} & 0\\
Y_c^{u(d)} & 0 & 0 \\
0 &  0 & Y_a^{u(d)} \\
\end{array}\right).
\label{Yuq}
\ee
All the Yukawa couplings are  real, because we assume that
CP is spontaneously broken.
So, the VEVs of the Higgs fields have to be complex
to obtain the CP phase of the CKM matrix.
It has been found  in \cite{Babu:2004tn} that for  successful predictions, i.e. sum rules among the fermion masses and
CKM parameters, 
 the scalar potential  should have an accidental $Z_2$ invariance
\be
h^{u,d}_+&=&\frac{1}{\sqrt{2}}
(h^{u,d}_1+h^{u,d}_2)
\rightarrow h^{u,d}_+~,~
h^{u,d}_- =\frac{1}{\sqrt{2}}
(h^{u,d}_1-h^{u,d}_2) \rightarrow -h^{u,d}_-.
\label{Hpm}
\ee
Then there will be  only nine independent parameters in the quark sector to
describe ten observables (six quark masses and four CKM parameters).

The superpotential for the Higgs sector consists of $\mu$ terms.
The only $Q_6$ invariant $\mu$ term is
$(H_1^u H_1^d+H_2^u H_2^d)$, and  no $H_3^u H_3^d$ and 
no mixing between the $Q_6$ doublet and singlet Higgs multiplets  are allowed.
Therefore, there is an accidental global $SU(2)$, implying
the existence of Nambu-Goldstone modes. 
In \cite{BK2010} the Higgs sector is extended to include 
a certain set of SM singlet Higgs multiplets to avoid 
this problem. It has been further found that
spontaneous breaking of the flavor symmetry $Q_6$ as well as CP invariance
can be realized without breaking
 the accidental  $Z_2$ invariance (\ref{Hpm}).
It has been also shown that, although the scale of  the singlet sector 
is of the same order as the heavy $SU(2)$ doublet Higgs bosons, 
an effective $\mu$ term $ W^{\rm eff} $  along with the soft-supersymmetry-breaking 
Lagrangian 
${\cal L}_{\rm soft}^{\rm eff}$ can well describe the original theory in that  sector,
where
\begin{eqnarray}
W^{\rm eff} &=&
\mu^{++}~(H^{u}_+ H^{d}_+ +H^{u}_- H^{d}_-)+
\mu^{+3}~H^{u}_+ H^{d}_3
+\mu^{3+}~H^{u}_3 H^{d}_+ ~,
\label{Weff}\\
{\cal L}_{\rm soft}^{\rm eff} &=&
m_{H^u}^2~(|h^{u}_+ |^2+|h^{u}_- |^2)+
m_{H^u_3}^2~|h^{u}_3|^2+
m_{H^d}^2~(|h^{d}_+ |^2+|h^{d}_- |^2)+
m_{H^d_3}^2~|h^{d}_3|^2\nonumber\\
& +&\left[~
B^{++}~(h^{u}_+ h^{d}_+ +h^{u}_- h^{d}_-)+
B^{+3}~h^{u}_+ h^{d}_3+
B^{3+}~h^{u}_3h^{d}_+ 
+h.c.\right]~.
\label{Leff}
\end{eqnarray}
(The notation $H_+^u$ etc should be self-evident, and the $A$ terms are suppressed.)
The parameters $\mu$'s and $B$'s  are complex,
which come from the complex VEVs of the singlet Higgs fields of the original theory.
So, the effective superpotential (\ref{Weff})  and the effective 
soft-supersymmetry-breaking Lagrangian (\ref{Leff}) break $Q_6$ and CP softly.
However, thanks to (\ref{Hpm}), 
the VEVs of the form
\be
<\phi_{-}^{u,d0}> &=& 0~,~ <\phi_{+}^{u,d0}>
 =\frac{v_{+}^{u,d}}{\sqrt{2}}\exp i \theta_+^{u,d}~, ~
<\phi_{3}^{u,d0}> 
=\frac{v_3^{u,d}}{\sqrt{2}}\exp i \theta_3^{u,d}
\label{vev1}
\ee
can be realized. (See (\ref{su2-comp}) for the notation.)

The CKM mixing matrix is given by
 \be
   V_{\rm CKM} &=& (U^{u}_L)^\dag  U_L^d=
 O^{uT}_L  P_q O^{d}_L ~,
\label{ckm}
 \ee
 where 
 \be
P_q  &=&\mbox{diag.}~(
1, \exp (i2\theta_q), \exp (i\theta_q))~,~
\theta_q =\theta^u_+-\theta^d_+-\theta^u_3
+\theta^d_3~, 
\label{phase}
\\
 O_{uL}   &\simeq & \left(
\begin{array}{ccc}
0.9992 & 0.04037 &  9.371\times 10^{-6} \\
0.04029 & -0.9974&  0.05978\\
-2.422\times 10^{-3} & 0.05973& 0.9982
\end{array}\right),\nn
\\
O_{uR} & \simeq   &\left(
\begin{array}{ccc}
-0.9992 &  0.04037&-9.025\times 10^{-5}\\
0.04029& 0.9973 &0.06207 \\
-2.515\times 10^{-3} &-0.06202 & 0.9981
\end{array}\right),
\label{UuR}
\ee
\be
O_{dL}  &\simeq& \left(
\begin{array}{ccc}
0.9760& -0.2176 &  -1.945 \times 10^{-3}\\
-0.2174& -0.9756 & 0.03116\\
8.679\times 10^{-3} &0.02999 &  0.9995
\end{array}\right),
\nn\\
O_{dR}  &\simeq& \left(
\begin{array}{ccc}
-0.9693 & -0.2460&  1.330\times 10^{-4}\\
-0.2190&0.8628 &  0.4557\\
0.1122&  -0.4416& 0.8901
\end{array}\right)~.\nn
\ee
The nine independent theory parameters are
$Y_a^{u,d} v_3^{u,d},
Y_c^{u,d} v_3^{u,d},Y_b^{u,d} v_+^{u,d},
Y_{b'}^{u,d} v_+^{u,d}$ and $\theta_q$, which
describe the CKM parameters and the quark masses as mentioned.
The set of the theory parameters is thus over-constrained.  Therefore,
there is not much freedom in the parameter space, 
and so it is sufficient to
consider a single point in the space of the theory parameters
of this sector. The orthogonal matrices (\ref{UuR}) are obtained for the following
Yukawa couplings:
\be
Y_a^{u} v_3^{u} &=&1.409 ~m_t ~,
Y_c^{u} v_3^{u}=2.135 \times10^{-4}  ~m_t ~,
Y_b^{u} v_+^{u}=0.0847 ~ m_t ~,
Y_{b'}^{u} v_+^{u}=0.0879 ~ m_t ~,\nn\\
Y_a^{d} v_3^{d} &=&1.258 ~m_b ~,
Y_c^{d} v_3^{d}=-6.037 \times
10^{-3} ~m_b~,Y_b^{d} v_+^{d}=0.0495  ~m_b ~,
Y_{b'}^{d} v_+^{u,d}=0.6447  ~m_b ~,\nn\\
\theta_q &=& -0.7125 ~.
\label{input}
\ee
With these parameter values we obtain \cite{Araki:2008rn}
\be
m_u/m_t &=& 0.609\times 10^{-5}~,~
m_c/m_t=3.73 \times 10^{-3}~,~
m_d/m_b=0.958 \times 10^{-3}~,\\
\label{ratio}
m_s/m_b & =& 1.69 \times 10^{-2}~,~
| V_{\rm CKM} |= 
\left( \begin{array}{ccc}
0.9740& 0.2266  & 0.00361
\\  0.2264   & 0.9731& 0.0414
  \\ 0.00858 &0.0407& 0.9991
\end{array}\right)~,\\
|V_{td}/V_{ts}| & = &0.211~,~
\sin 2\beta (\phi_1) = 0.695~,~\bar{\rho}=0.152~,~\bar{\eta}=0.343~.
\label{ckm-parameters}
\ee
The mass ratio (\ref{ratio}) is defined at $M_Z$ and consistent with the recent up-dates
of \cite{Xing:2007fb}, and  the CKM parameters above  agree very well with those of
Particle Data Group \cite{Amsler:2008zzb} and CKM fitter groups \cite{CKMfitter,UTfit}.
(See \cite{Kubo:2003iw} for the prediction of the model in the lepton sector.)

So far we have discussed only the Yukawa sector.
To compute the one-loop corrections to $\delta_{LL}$'s, we need to fix the Higgs sector.
It is convenient to make a phase rotation of the Higgs superfields
so that their VEVs become real:
\begin{eqnarray}
\tilde{H}^{u,d}_{\pm}  &=& H^{u,d}_{\pm} e^{-i \phi^{u,d}_{+}},~
\tilde{H}^{u,d}_{3} = H^{u,d}_{3} e^{-i \phi^{u,d}_{3}}.
\label{Htilde}
\end{eqnarray}
Then we define
\begin{equation}
\left( \begin{array}{c}
\Phi^{u,d}_L \\\Phi^{u,d}_H \\ \Phi_-^{u,d}
\end{array}
\right)
:= 
\left( \begin{array}{ccc}
 \cos \gamma^{u,d} & \sin \gamma^{u,d} & 0
\\
-\sin \gamma^{u,d} & \cos \gamma^{u,d} & 0
\\
0              & 0             & 1
\end{array}
\right)
\cdot
\left( \begin{array}{c}
\tilde{H}_3^{u,d} \\ \tilde{H}_+^{u,d} \\ \tilde{H}^{u,d}_{-}
\end{array}
\right)
\end{equation}
where
\begin{eqnarray}
\cos\gamma^{u,d} &=& v_3^{u,d}/v^{u,d}~,~
\sin\gamma^{u,d} = v_+^{u,d}/v^{u,d}~,~
v^{u,d}=\sqrt{(v_3^{u})^2+(v_+^{u})^2})~.
\label{cosgamma}
\end{eqnarray}
The components of the $SU(2)$ doublet Higgs multiplets are defined

\begin{eqnarray}
\Phi^u_I &=&\left( \begin{array}{c}
\Phi^{u+}_I\\
\Phi^{u0}_I
\end{array}\right),~
\Phi^d_I =\left( \begin{array}{c}
\Phi^{d0}_I\\
\Phi^{d-}_I
\end{array}\right),~I=L,H,-.
\label{doublet}
\end{eqnarray}
 The light and heavy MSSM-like Higgs scalars  are then given by
\begin{eqnarray}
( v + h-i X )/\sqrt{2} &=&( \phi^{d0}_L)^*\cos\beta
+ (\phi^{u0}_L)\sin\beta~,\label{hX}\nn\\
(H+i A)/\sqrt{2} &=&- (\phi^{d0}_L)^*\sin\beta
+ (\phi^{u0}_L)\cos\beta~,\label{HA}\label{HP}\\
G^+ = - (\phi^{d-}_L)^*\cos\beta &+ & ( \phi^{u+}_L)\sin\beta~,~
H^+ =
 (\phi^{d-}_L)^*\sin\beta+\phi^{u}_L\cos\beta~,\nn
 \label{CH}
\end{eqnarray}
where 
$X$ and $G^+$ are the Nambu-Goldstone fields, 
$\phi$'s are  scalar components of (\ref{doublet}), and
$v =\sqrt{v_u^2+v_d^2}
~(\simeq 246$ GeV) and $\tan\beta=v_u/v_d$.

\subsection{Calculation of $\delta$'s}
To apply the general formula (\ref{dLR-mu}), (\ref{dLL-quart}), (\ref{dLL-cs}) and (\ref{dLL-f}),
we have to
compute the mass matrix for  the charged Higgs bosons ${\bf M}_c^2$ and 
fermions ${\bf M}_h^F$.
Note that since the $Z_2$ invariance is unbroken, $Z_2$ even and odd fields do not mix
with each other.
We find that the mass matrix of the charged $Z_2$ even Higgs bosons 
has the form
\begin{eqnarray}
{\bf M}^2_{c,\rm even} &=&\left( \begin{array}{ccc}
\frac{2B_L}{s_{2\beta}}+c_W^2 M_Z^2 & -\hat{m}^2_{uLH}/c_\beta &
 -\hat{m}^{2*}_{dLH}/s_\beta \\
 -\hat{m}^{2*}_{uLH}/c_\beta & -\hat{m}_{uH}^2 
 - c_{2\beta}s_W^2M_Z^2 &
 B_{H}^* \\
  -\hat{m}^2_{LH}/s_\beta &  B_{H}&
   -\hat{m}_{dH}^2 
 +c_{2\beta} s_W^2 M_Z^2 
\end{array}\right)
\label{mce}
\end{eqnarray}
in the basis of $(H^+,~\phi_H^{u+},~(\phi_H^{d-})^*)$, 
and that of the
the charged $Z_2$ odd Higgs bosons is 
\begin{eqnarray}
{\bf M}^2_{c,\rm odd} &=&\left( \begin{array}{cc}
-m_{H^u}^2+|\mu^{++}|^2 + c_{2\beta}c_W^2 M_Z^2/2  & B^{++*}\\
B^{++} &
-m_{H^d}^2+|\mu^{++}|^2 - c_{2\beta}c_W^2 M_Z^2/2
 \end{array}\right)
 \label{mco}
\end{eqnarray}
in the $(\phi_-^{u+},~(\phi_-^{d-})^*)$ basis,
where $c_W=\cos \theta_W=(1-0.23)^{1/2}$,  $s_{2\beta}=\sin 2\beta$ etc.
The mass parameters in (\ref{mce})  and (\ref{mco}) are defined as
\begin{eqnarray}
-\hat{m}_{u(d)LH}^2 &=& -
(c_{\gamma^{u,d}}s_{\gamma^{u,d}} m_{H^{u(d)}}^2-m_{H^{u(d)}_3}^2)
+\mu_{H}\mu_{LH(HL)}^*
+\mu_{HL(LH)}\mu_{L}^*~,\nn\\
-\hat{m}_{u(d)H}^2 &=&-
(c_{\gamma^{u(d)}}^2 m_{H^{u(d)}}^2+
s_{\gamma^{u(d)}}^2 m_{H^{u(d)}_3}^2)
 +|\mu_{H}|^2
+|\mu_{HL(LH)}|^2~,\nn\\
B_L &=& s_{\gamma^u} s_{\gamma^d}B^{++}e^{i(\theta_+^u+\theta_+^d)}+
s_{\gamma^u} c_{\gamma^d}B^{+3}e^{i(\theta_+^u+\theta_3^d)}+
c_{\gamma^u} s_{\gamma^d}B^{3+} e^{i(\theta_3^u+\theta_+^d)}~,
\label{mus}\\
B_H &=&c_{\gamma^u} c_{\gamma^d}B^{++}e^{i(\theta_+^u+\theta_+^d)} - 
c_{\gamma^u} s_{\gamma^d}B^{+3} e^{i(\theta_+^u+\theta_3^d)}- 
s_{\gamma^u} c_{\gamma^d}B^{3+}e^{i(\theta_3^u+\theta_+^d)}~,
\nn\\
\mu_{HL} &=&c_{\gamma^u} s_{\gamma^d}\mu^{++}e^{i(\theta_+^u+\theta_+^d)}+
c_{\gamma^u} c_{\gamma^d}\mu^{+3}e^{i(\theta_+^u+\theta_3^d)} + 
-s_{\gamma^u} s_{\gamma^d}\mu^{3+}e^{i(\theta_3^u+\theta_+^d)}~,\nn\\
\mu_{LH} &=&
s_{\gamma^u} c_{\gamma^d}\mu^{++}e^{i(\theta_+^u+\theta_+^d)}
- s_{\gamma^u} s_{\gamma^d}\mu^{+3}e^{i(\theta_+^u+\theta_3^d)}+
c_{\gamma^u} c_{\gamma^d}\mu^{3+}e^{i(\theta_3^u+\theta_+^d)}~,\nn\\
\mu_{L}&=&
s_{\gamma^u} s_{\gamma^d}\mu^{++}e^{i(\theta_+^u+\theta_+^d)}+
s_{\gamma^u} c_{\gamma^d}\mu^{+3}e^{i(\theta_+^u+\theta_3^d)}+
c_{\gamma^u} s_{\gamma^d}\mu^{3+}e^{i(\theta_3^u+\theta_+^d)}~,\nn\\
\mu_H &=& c_{\gamma^u} c_{\gamma^d}\mu^{++}e^{i(\theta_+^u+\theta_+^d)} - 
c_{\gamma^u} s_{\gamma^d}\mu^{+3} e^{i(\theta_+^u+\theta_3^d)}- 
s_{\gamma^u} c_{\gamma^d}\mu^{3+}e^{i(\theta_3^u+\theta_+^d)}~.\nn
\end{eqnarray}
For the charginos  we find
\be
{\bf M}_{h,\rm even}^F & = &\left( \begin{array}{ccc}
M_2 & \sqrt{2}c_W s_\beta M_Z &0 \\
 \sqrt{2}c_W c_\beta M_Z &\mu_L & \mu_{HL} \\
0 &\mu_{LH} & \mu_H \\
 \end{array}\right),~M^F_{h,\rm odd}=\mu^{++}~.
 \label{mhf}
\end{eqnarray}
 
\begin{table}[ht]
\vspace*{-0.05in}
$$\begin{array}{|c|c||c|c|}\hline
\tan\gamma^u & -0.1188 & \tan\gamma^d  & -0.9480 \\\hline
\tan\beta &3.180 & &  \\
\hline
 &[\mbox{TeV}] &   &[ \mbox{TeV}^2] \\
 \hline
\mu^{++} e^{i(\theta_+^u+\theta_+^d)} & 
(0.900+ i ~0.034)
& B^{++} e^{i(\theta_+^u+\theta_+^d)}& 
(4.179^2+i~ 0.7094^2)\\ \hline
\mu^{+3}e^{i(\theta_+^u+\theta_3^d)} &  
(0.230+i ~0.120 )
& B^{+3} e^{i(\theta_+^u+\theta_3^d)} &
(2.835^2+i~ 0.4286^2)\\ \hline
\mu^{3+} e^{i(\theta_3^u+\theta_+^d)}& 
(-0.660+i~ 0.050 )
& B^{3+}  e^{i(\theta_3^u+\theta_+^d)}& 
(-3.106^2+i~ 0.1918^2)\\
\hline
   &[ \mbox{TeV}^2] &   & [ \mbox{TeV}^2] \\ \hline
m_{H^u}^2   &  -6.457^2 
   &    m_{H^d}^2    &  -7.265^2\\ \hline
m_{H^u_3}^2&  -1.261^2  & m_{H^d_3}^2  &  -2.071^2 \\ 
\hline
\end{array}$$
\caption{ \footnotesize{A representative set of parameter values.
The phases are not fixed, except for $\theta_q$ (see (\ref{input})
and (\ref{phase}).)  }}
\label{example}
\end{table}
To explicitly calculate $\delta$'s, 
we consider a representative set of parameter values
 which is given in Table \ref{example}.
For the parameters given in Table \ref{example}  we find the Higgs mass spectrum:
 \begin{eqnarray}
{\bf M}_{c, \rm even}& :&\left(~8.25, 5.69, 1.61~\right)~~
\mbox{TeV}~,~{\bf M}_{c,\rm odd} ~:~\left(~8.15, 5.45~\right)~~
\mbox{TeV}~,\nn\\
{\bf M}_{h,\rm even}^F&: &\left(~1.14, ~0.506,
~0.144~\right)~~
\mbox{TeV}~,~
M_{h,\rm odd}^F = 0.900~\mbox{TeV}~.
\label{h-mass}
\end{eqnarray}
The lightest one of  ${\bf M}_{c, \rm even}$ and two of 
${\bf M}_{h,\rm even}^F$ correspond to
 the charged Higgs boson and fermions of the MSSM.
 Note that the charged fermions are much lighter than the charged bosons,
because the EDM constraints require small $\mu$'s 
as we see from (\ref{dLR-mu}) \cite{Babu:2009nn}.
 We have chosen the parameter values in this way, because we have to suppress
 three-level FCNCs as well as the EDMs.

The $A$ terms (which are  suppressed 
in the soft-supersymmetry-breaking Lagrangian (\ref{Leff}) ) and
soft scalar mass terms  have the same family symmetry 
as  the Yukawa sector even in the effective theory \cite{BK2010}.
Consequently, the
soft scalar mass matrices have the following form:
\be
{\bf \tilde{m}^2}_{aLL}& =&
{m}^2_{\tilde{a}}~ \mbox{diag.}~
(a_{L}^{a}~,~ a_{L}^{a}~,~ b_{L}^{a})~~(a=q,l)~,\nn\\
{\bf \tilde{m}^2}_{aRR}& =&
{m}^2_{\tilde{a}}~ \mbox{diag.}~
(a_{R}^{a}~,~ a_{R}^{a}~,~ b_{R}^{a})~~(a=u,d,e)~,
\label{scalarmass}\\
\left({\bf \tilde{m}^2}_{aLR}\right)_{ij} 
&=&
A_{ij}^a\left( {\bf m}^a \right)_{ij}=
\tilde{A}_{ij}^a ~m_{\tilde{a}}~\left( {\bf m}^a \right)_{ij} 
~~(a=u,d,e)~,\nn
\ee
where ${m}_{\tilde{a}}$ denote the average of the  squark 
and slepton masses, respectively,   $(a_{L(R)}^a, b_{L(R)}^a)$ are
dimensionless free real parameters, 
 $A_{ij}^{a}$ are free parameters of dimension one,
 and ${\bf m}^a$ are  the respective fermion mass matrices.
According to \cite{Hall:1985dx,Gabbiani:1988rb}
we define the tree-level  supersymmetry-breaking soft mass insertions as
\be
\delta_{LL(RR)}^{a0} &=&
U_{aL(R)}^{\dagger} ~{\bf \tilde{m}^2}_{aLL(RR)}~
 U_{aL(R)}/{m}^2_{\tilde{a}}~,~
\delta_{LR}^{a0} =
U_{aL}^{\dagger}~ 
{\bf \tilde{m}^2}_{aLR} ~U_{aR}/{m}^2_{\tilde{a}} 
\label{Delta1}
\ee
in the super CKM basis. Only the $A$ term contributions are included into
the left-right mass matrices ${\bf \tilde{m}^2}_{aLR}$ in (\ref{scalarmass}), and 
the $\mu$ term contributions (\ref{dLR-mu}) will be added to $\delta_{LR}$ separately.
Note that $a_{L,R}^a$ and $A_{ij}^a$ are all real,
because of  CP invariance of the original theory, and that
the structure (\ref{scalarmass}) is the consequence of the flavor symmetry $Q_6$.
Since $Q_6$ is only spontaneously broken in the original theory and only softly broken
by the $\mu$ terms and $B$ terms in the effective theory described by
(\ref{Weff}) with (\ref{Leff}), 
there will be no divergent contributions to the non-diagonal elements
of $\delta_{RR}$ and $\delta_{LL}$. 
We have explicitly checked
the cancellation of the divergences up to terms proportional to the square of
the quark masses times gauge coupling squared, which we have anyway neglected.
Since these terms are partially included in the following
calculations, the cancellation of the renormalization
scale $Q$ dependence is not exact. 
We find about $0.4 \%$ change of the non-diagonal elements of
$\delta_{LL}$ against the change of
$Q $ by two orders of
magnitude.  In the following calculations we set $Q$ equal to $m_{\tilde{d}}$.

We find
from (\ref{dLL-quart}), (\ref{dLL-cs}) and (\ref{dLL-f})
\be
(\delta^{d}_{12})_{LL}
&=&  (\delta^{d}_{21})_{LL}^*\simeq 
-2.6 \times 10^{-4}~\Delta a_L^{q}+
(1.2\times 10^{-3}- i~1.2\times 10^{-6})
\left[\frac{0.5~\mbox{TeV}}{m_{\tilde{d}}}\right]^2~,
\nn\\
(\delta^{d}_{13})_{LL}
&=& (\delta^d_{31})_{LL}^*
\simeq  -8.7\times 10^{-3}~\Delta a_L^q+
(0.50- i~1.9)\times 10^{-2}
\left[\frac{0.5~\mbox{TeV}}{m_{\tilde{d}}}\right]^2~,
\nn\\
(\delta^{d}_{23})_{LL}
&=& (\delta^{d}_{32})_{LL}^*
\simeq  -3.0 \times 10^{-2}~\Delta a_L^{q}
-(0.28+ i~8.7)\times 10 ^{-2}
\left[\frac{0.5~\mbox{TeV}}{ m_{\tilde{d}}}\right]^2~,
\label{deltaLL-d}\\
(\delta^{d}_{12})_{RR}
&=& (\delta^d_{21})_{RR}^*
\simeq  5.0 \times 10^{-2} ~\Delta a_R^{d} ~,~
(\delta^{d}_{13})_{RR}
= (\delta^d_{31})_{RR}^*
\simeq -0.10~\Delta a_R^{d}~, \nn\\
(\delta^{d}_{23})_{RR}
&=& (\delta^d_{32})_{RR}^*
\simeq  0.39~\Delta a_R^{d}~,\nn\\
\Delta a_{L}^{q} &=& a_{L}^{q}-b_{L}^{q}~,~
\Delta a_{R}^{d} =a_{R}^{d}-b_{R}^{d}~,
\ee
where terms proportional to $\Delta a_L^{q}$ and $ \Delta a_R^{d}$
are the tree-level insertions. We see that the one-loop effects,
their real as well as their imaginary parts,
to  $\delta_{LL}$ are comparable
to the tree-level ones,  while the imaginary part of the $(1,2)$ element of the one-loop effect
is much smaller than its real part. This is a good news, because CP violation in
the first generation of quarks is very small, while 
 CP violation in
the third generation may be large. This is a consequence of the hierarchical structure
of the Yukawa couplings as one can see from (\ref{Yuq}) and (\ref{input}).
We have not included the
one-loop corrections to $\delta_{RR}$, because due the smallness
of the Yukawa couplings in the down-quark sector they are
very small compared with the tree-level contributions.
For the left-right insertions  we find
\be
(\delta^{{d}}_{12})_{LR}
&\simeq &
1.9 (\tilde{A}_{1}^{d}-\tilde{A}_{2}^{d})\times 10^{-5} 
\left[\frac{0.5 ~\mbox{TeV}}{m_{\tilde{d}} }\right]+
(2.9- i~0.11)\times 10^{-4}
\left[\frac{0.5~\mbox{TeV}}{ m_{\tilde{d}}}\right]^2~,
\nn\\
(\delta^{{d}}_{21})_{LR}
&\simeq &
(- 2.2\tilde{A}_{1}^{d}+1.7  \tilde{A}_{2}^{d} )\times 10^{-5}
\left[\frac{0.5 ~\mbox{TeV}}{m_{\tilde{d}} }\right]-
(2.9- i~0.11)\times 10^{-4}
\left[\frac{0.5~\mbox{TeV}}{ m_{\tilde{d}}}\right]^2~,
\nn\\
(\delta^{{d}}_{13})_{LR}
&\simeq &
(1.0 \tilde{A}_{1}^{'d}+4.0 \tilde{A}_{2}^{'d}) \times 10^{-5}
\left[ \frac{0.5 ~\mbox{TeV}}{m_{\tilde{d}} }\right]-
(2.1- i~0.09)\times 10^{-4}
\left[\frac{0.5~\mbox{TeV}}{ m_{\tilde{d}}}\right]^2~,\nn\\
(\delta^{{d}}_{31})_{LR}
&\simeq &
5.8 \tilde{A}_{2}^{d} \times 10^{-4}
\left[ \frac{0.5 ~\mbox{TeV}}{m_{\tilde{d}} }\right]-
(4.2 - i~0.17)\times 10^{-3}
\left[\frac{0.5~\mbox{TeV}}{ m_{\tilde{d}}}\right]^2~,
\label{deltaLR-d}\\
(\delta^{{d}}_{23})_{LR}
&\simeq &
1.4 \tilde{A}_{2}^{'d}\times 10^{-4}
\left[ \frac{0.5 ~\mbox{TeV}}{m_{\tilde{d}} }\right]-
(1.0 - i~0.04)\times 10^{-3}
\left[\frac{0.5~\mbox{TeV}}{ m_{\tilde{d}}}\right]^2~,\nn\\
(\delta^{{d}}_{32})_{LR}
&\simeq &
-2.3 \tilde{A}_{2}^{d} \times 10^{-2} 
\left[ \frac{0.5 ~\mbox{TeV}}{m_{\tilde{d}} }\right]+
(1.7 - i~0.07)\times 10^{-2}
\left[\frac{0.5~\mbox{TeV}}{ m_{\tilde{d}}}\right]^2~,\nn
\ee
where the first terms come from the $A$ terms, and the second ones are from
the $\mu$ terms (\ref{dLR-mu}), and $\tilde{A}^d$'s in 
(\ref{deltaLR-d}) are dimensionless  free parameters.
The imaginary part of the $(1,1)$ element of $\delta_{LR}$ is strongly constraint
by EDMs \cite{Gabbiani:1996hi,Abel:2001vy}.  The $A$-term contributions are real 
(because of the CP invariance of the original theory), while
the $\mu$ term contributions are complex (because spontaneous CP violation generates
complex $\mu$ terms). We find, using (\ref{dLR-mu}), 
\be
\left\{ \begin{array}{c}
\mbox{Im} (\delta^{{d}}_{11})_{LR} = 3.7 \\
\mbox{Im} (\delta^{{u}}_{11})_{LR} = 1.3 \end{array}
~\times 10^{-6} \times \left[\frac{0.5~\mbox{TeV}}{ m_{\tilde{d}}}\right]^2~,
\right.
\label{edm}
\ee
which are of the order of the upper bound \cite{Gabbiani:1996hi}.

\subsection{Mixing of the neutral mesons and new CP phases}
We now apply our results (\ref{deltaLL-d}) and (\ref{deltaLR-d})  
to the  mixing of the neutral meson systems.
Here we assume that the tree-level contributions to the mixing coming from the heavy
neutral Higgs boson exchange are small. In \cite{Kifune:2007fj,Kawashima:2009jv,Babu:2009nn} 
it has been found that if their masses
 are larger than  several  TeV in the present model, then FCNCs and CP are suppressed. In the present case
 with the parameters given in Table \ref{example}, 
 the mass of the lightest flavor-changing  neutral 
 Higgs boson is $5.7$ TeV, and so the assumption may be justified.
The total matrix element  $M_{12}^q$ in the neutral meson mixing can be written as
\be
M_{12}^q &= &M^{SM,q}_{12} + M^{ SUSY,q}_{12}~,
\ee
where $M^{SM,q}_{12}$ is the SM contribution, and 
$M^{ SUSY,q}_{12}$ is the SUSY contribution, whose dominant contribution is given by
\cite{Gabbiani:1996hi}
(see e.g. \cite{Gorbahn:2009pp} for a more refined calculation)
\be
 M^{ SUSY,s}_{12} &=&
-\frac{\alpha_S^2}{324 m_{\tilde{d}}^2}M_s f_{B_s}^2~B_s\left\{~
[~(\delta_{32}^d)^2_{LL}+(\delta_{32}^d)^2_{RR}~][~24 x f_6(x)+66 \tilde{f}_6(x)~]
\right.\nn\\
& &\left.
+(\delta_{32}^d)_{LL}(\delta_{32}^d)_{RR}
\left[\left( 384 ~R_s+72\right) x f_6(x)-
\left( 24  R_s -36\right) \tilde{f}_6(x)\right]
\right.\nn\\
 & &\left.
-132~ [~(\delta_{32}^d)_{LR}^2 +(\delta_{23}^d)_{LR}^{*2}~] 
 R_s x f_6(x)\right.\nn\\
& &\left.
-(\delta_{32}^d)_{LR}(\delta_{23}^d)_{LR}^*
\left[ ~144 ~ R_s +84~\right] \tilde{f}_6(x)~ \right\}~,
\label{m12-susy}
\ee
where
\be
R_s &= &\left(~\frac{M_s}{m_s+m_d}~\right)^2~,\nn\\
f_6(x) &=& \frac{6(1 + 3x)\ln x + x^3 - 9 x^2 - 9 x + 17}{6 (x - 1)^5}~,
\label{f6}\\
\tilde{f}_6(x) &=& \frac{6x(1 + x)\ln x - x^3 - 9 x^2 + 9 x + 1}{3 (x - 1)^5}
~~\mbox{with}~~x=m_{\tilde{g}}/m_{\tilde{d}}~,\nn
\ee
similarly for $K$ and $B_d$, and $m_{\tilde{g}}$ is the 
gluino mass.  For the calculations below we assume that
the bag parameters $B_K, B_d, B_s$ are one, and $\alpha_S=0.12$.
The other parameters 
are given in Table \ref{input2}. Since (\ref{m12-susy}) is a one-loop result,
the one-loop effect to the soft mass insertions in fact means
a two-loop effect    like Fig.~3
\begin{figure}[htb]
\label{twoloop}
\begin{center}\begin{picture}(300,110)(0,45)
\ArrowLine(80,40)(110,40)
\DashArrowLine(110,40)(143,40){3}
\DashArrowLine(138,40)(170,40){3}
\ArrowLine(170,40)(200,40)
\ArrowLine(110,100)(80,100)
\DashArrowLine(143,100)(110,100){3}
\DashArrowLine(170,100)(138,100){3}
\ArrowLine(200,100)(170,100)
\ArrowLine(110,100)(110,40)
\ArrowLine(170,40)(170,100)
\DashArrowArc(140,80)(20,90,450){3}
\Text(95,25)[b]{$s_R$}
\Text(125,25)[b]{$\tilde{s}_R$}
\Text(155,25)[b]{$\tilde{b}_R$}
\Text(185,25)[b]{$b_R$}
\Text(140,37)[b]{$\times$}
\Text(100,70)[b]{$\tilde{g}$}
\Text(180,70)[b]{$\tilde{g}$}
\Text(95,110)[b]{$b_L$}
\Text(125,110)[b]{$\tilde{b}_L$}
\Text(155,110)[b]{$\tilde{s}_L$}
\Text(185,110)[b]{$s_L$}
\Text(140,70)[b]{$\phi^{u+}$}
\end{picture}\end{center}
\caption[]{An example of two-loop contribution to 
$M^{ SUSY,s}_{12}$. One-loop contribution to
the insertion $(\delta^d_{32})_{LL}$, the $\phi^{u+}$ loop in the box,
means a two-loop effect on $M^{ SUSY,s}_{12}$.}
\end{figure}
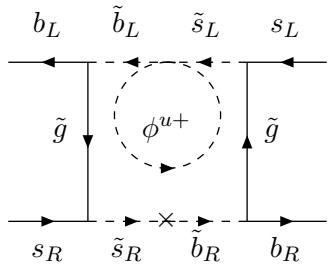

We follow \cite{LN2006} to parameterize new physics effects as
(see (\ref{M12allg}))
\be
M^{ SM,q}_{12} + M^{ SUSY,q}_{12}
&=&  M^{ SM,q}_{12} \cdot \Delta_q\, ,
\label{M12}
\ee
and 
consider the observables
$\Delta M_{q}$, $\Delta \Gamma_{q}$ and the flavor specific CP-asymmetry 
$a_{sl}^{q}$ in terms of the complex number 
$\Delta_{q} = |\Delta_{q}| e^{i \phi^{\Delta}_{q}}$, where $q=d,s$, and
\be
\Delta M_q  & = & 2| M^{ SM,q}_{12}| \cdot |\Delta_q | ~,~
\Delta \Gamma_q  = 2 |\Gamma^q_{12} |
     \, \cos \left( \phi_q^{ SM} + \phi^\Delta_q \right)~,
\nn\\
a_{sl}^q
& =&
 \frac{| \Gamma^q_{12} |}{| M^{ SM,q}_{12} |}
\cdot \frac{\sin \left( \phi_q^{ SM} + \phi^\Delta_q\right)}{|\Delta_q|}~.
\label{afs} 
\ee
The SM values are given e.g.  in \cite{LN2006}, in which the results of
\cite{BBD06,BBGLN98,BBGLN02,BBLN03,rome03} are used:
\be
2~M^{{ SM},d}_{12}&=&
0.56(1\pm 0.45)\exp (i 0.77)
~\mbox{ps}^{-1}~,\nn\\
 2~M^{{ SM},s}_{12} &=&
20.1(1\pm 0.40)\exp (-i 0.035)
~\mbox{ps}^{-1}~,
\nn\\
 \Delta \Gamma_d^{ SM}  &=&
 (26.7 ~{}^{+5.8}_{-6.5}) \times10^{-4}~\mbox{ps}^{-1}~,~
 \Delta \Gamma_s^{ SM}  =
 0.096\pm0.039 ~\mbox{ps}^{-1}~,\nn\\
\ee
where the errors are dominated by the uncertainty in the 
decay constants and bag parameters.
\vspace{0.5cm}
\begin{table}
\begin{center}
\begin{tabular}{|c|c||c|c|}
\hline
Input &  &  Input &   \\ \hline
$f_K$ & $(159.8\pm 1.4 \pm 0.44)\times 10^{-3}$ GeV & 
 $f_{B_d}$ & $0.194\pm 0.032$ GeV  \\ \hline
 $f_{B_s}$ & $0.240\pm 0.040$ GeV  &  & 
\\ \hline
$M_K$ & $0.497648
\pm 0.000022$ GeV & $\Delta M_K^{\rm exp}$ & $(0.5292\pm
0.0009)\times 10^{-2}~\mbox{ps}^{-1}$ 
\\ \hline
$M_{s}$ & $5.3661\pm 0.0006$ GeV
 & $\Delta M_{s}^{\rm exp}$ & $17.77\pm 0.10\pm 0.07
 ~\mbox{ps}^{-1}$
\\ \hline
$M_{d}$ & $5.27950\pm 0.00033$ GeV 
& $\Delta M_{d}^{\rm exp}$ & 
$0.507\pm 0.005 ~\mbox{ps}^{-1}$ 
\\ \hline
$m_d(2 \mbox{GeV})$ & $(5.04 ~{}^{+0.96}_{-1.54})\times 10^{-3}$ GeV 
& $m_s (2 \mbox{GeV})$ & $0.105 ~{}^{+0.025}_{-0.035}$ GeV
\\ \hline
$m_d(m_b)$ & $(4.23~{}^{+1.74}_{-1.71})
\times 10^{-3}$ GeV & $m_s (m_b)$ & $0.080\pm 0.022$ GeV
\\ \hline
$m_b(m_b)$ & $4.20\pm 0.07$ GeV & &
\\ \hline
\end{tabular}
\caption{ \footnotesize{Parameter values used in the text
(see also Ref.~\cite{Bona:2006ah}).
For the calculations in the text we use only the central values.
$f_K, M_{K,d,s}, \Delta M_{K,d,s}^{\rm exp}$ are from 
\cite{Amsler:2008zzb}.
$f_{B_s}$
belongs to the conservative sets of \cite{LN2006}
(see the references therein), and 
$f_{B_d}$ is obtained from $f_{B_s}/\xi$ with $\xi=1.24$.
$m_d (2 \mbox{GeV})$ and $m_s (2 \mbox{GeV})$ are from 
\cite{Amsler:2008zzb}, while those at $m_b$ are taken from \cite{Xing:2007fb}. }}
\label{input2}
\end{center}
\end{table}

We use the central values of (\ref{phid}),  (\ref{phis}) and Table \ref{input2} for our calculations,
while requiring the constraints 
\be
\left. \begin{array}{c}0.6 \\
0.8 \end{array} \right\} & < & \frac{\Delta M_{d,s}}{\Delta M_{d,s}^{\rm exp }}  < 
\left\{ \begin{array}{c}1.4 \\ 1.2 \end{array}~,~
\frac{2 | M^{ SUSY,K}_{12} |}{\Delta M_{K}^{\rm exp }} <
\left\{\begin{array}{c} 2\\1 \end{array}~~\begin{array}{c} I \\ II \end{array}~
\right.\right.
\label{const1}
\ee
and
\be
\frac{\mbox{Im} M^{ SUSY,K}_{12}}{\sqrt{2} \Delta M_K^{\rm exp }}
  & <  & \epsilon_K=2.2\times 10^{-3}~,
\label{const2}
\ee
where I and II are a conservative and  an optimistic
set of constraints, respectively.

As we can see from (\ref{deltaLR-d}), except for $(\delta_{23,32}^d)_{LR}$, the $\mu$ term contributions 
(the second terms) are larger than the $A$ term contributions by 
an order  of magnitude  (if $A$'s are of $0(1)$). Therefore, we include only 
the $\mu$ term contributions to $M^{ SUSY,K}_{12}$ 
and $M^{ SUSY,d}_{12}$.
Further, $(\delta_{23,32}^d)_{LR}$ are  constrained by $b\to s \gamma$ 
\cite{Bertolini:1990if,Gabbiani:1996hi,Misiak:1997ei}, and
have to satisfy $|(\delta_{23,32}^d)_{LR}| \lsim  10^{-2}$. So, 
$|(\delta_{23,32}^d)_{LR}| $ are saturated with the $\mu$ term contribution.
We could choose a positive $O(1)$ value for $\tilde{A}_2^d$ so that
the $A$ term contribution cancels the $\mu$ term contribution, but
the effect on $|(\delta_{23}^d)_{LR}|$ is negligibly  small.
We therefore neglect the $A$ term contribution  in $(\delta_{23,32}^d)_{LR}$, too.
Under this situation, as we can see from (\ref{deltaLL-d}), (\ref{deltaLR-d})
and (\ref{m12-susy}),  given
 the set of parameters of Table \ref{example} and \ref{input2} with 
 $x=m_{\tilde{g}}/m_{\tilde{d}}=1$, 
the free parameters are only $\Delta a_{L}^d$
and $\Delta a_{R}^d$. So, it is absolutely non-trivial to satisfy  the  constraints
(\ref{const1}) and (\ref{const2}) while having a large CP violation in the $B^0$ mixing.

\begin{figure}[htb]
\includegraphics*[width=0.6\textwidth]{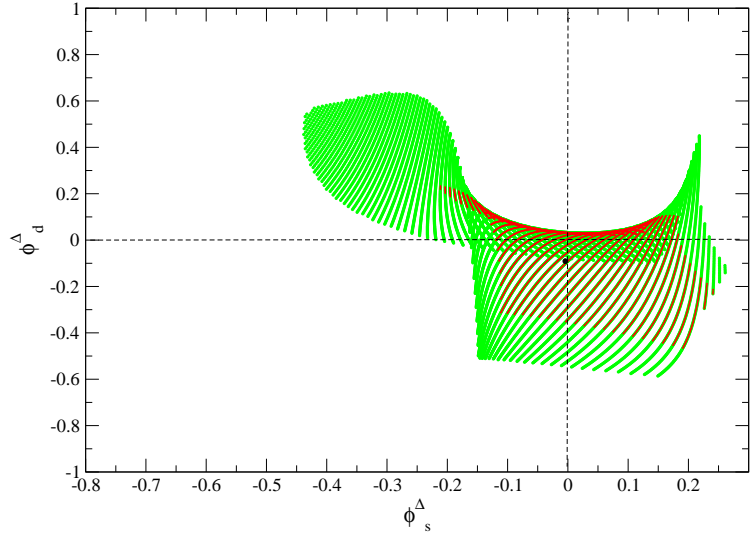}
\caption{\label{dphis-dphid}\footnotesize
The theoretical values in the  $\phi_s^\Delta-\phi_d^\Delta$ plane
for the  parameter values  given in Tables \ref{example} and 
\ref{input2}. The constraint
I  (II) of (\ref{const1}) with (\ref{const2}) is satisfied by the green (red) region. 
The  $\bullet$ is the SM value.}
\end{figure}

\begin{figure}[htb]
\includegraphics*[width=0.6\textwidth]{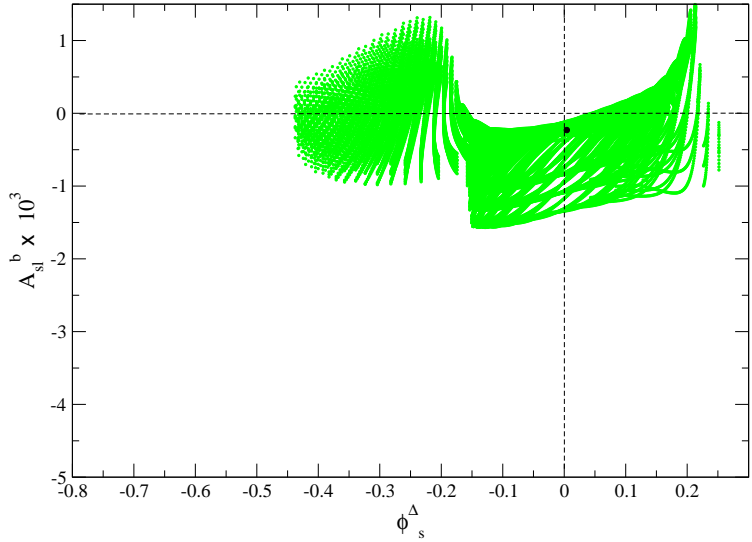}
\caption{\label{dphis-Asl}\footnotesize
The theoretical values  in the  $\phi_s^\Delta-A_{sl}^b$ plane
for the  same parameter values as Fig.~\ref{dphis-dphid}. 
We impose only the constraint
I   with (\ref{const2}). 
The  SM value $\bullet$ is obtained from  (\ref{asl-sm}) and (\ref{phis}).
The recent experimental values are, respectively,
$A_{sl}^b = -(9.57 \pm 2.51 \pm 1.46) \cdot 10^{-3}~$ \cite{Abazov:2010hv} and
$\phi_s^\Delta \in [-0.2,-2.8]~$\cite{combi}
($\phi_s^\Delta \in [0,-1]~$\cite{CDF}).}
\end{figure}

In Fig.~\ref{dphis-dphid} we plot $\phi_d^\Delta$ against $\phi_s^\Delta$
for the  parameter values  given in Tables \ref{example} and 
\ref{input2}. The green (red) region satisfies the constrains
I  (II) with (\ref{const2}).
The  SM value, (\ref{phid}) and (\ref{phis}), is denoted by $\bullet$.
Since no large CP phase can be generated
without  the loop effects of the 
extra heavy Higgs bosons in this model \cite{Kawashima:2009jv,Babu:2009nn}, 
the large $\phi_{d,s}^\Delta$
in the  predicted region of  Fig.~\ref{dphis-dphid} 
is entirely due the loop effects.
Fig.~ \ref{dphis-Asl}  shows the theoretical values in the 
$\phi_{s}^\Delta-A_{sl}^b$ plane, where
the same sign dimuon asymmetry $A_{sl}^b$ is computed  from (\ref{Asl}).
Here we have imposed only the conservative constraint I 
of (\ref{const1}) with (\ref{const2}), and assumed 
that 
$| \Gamma^{d}_{12} |/| M^{{ SM},d}_{12} |
=(52.6~{}^{+11.5}_{-12.8})\times 10^{-4}~,~
| \Gamma^{s}_{12} |/| M^{{ SM},s}_{12} |
=(4.97\pm 0.94)\times 10^{-3}$ \cite{LN2006}.
The  SM value $\bullet$ is obtained from  (\ref{asl-sm}) and (\ref{phis}).
The theoretical values in Fig.~ \ref{dphis-Asl} should be compared with
the recent experimental measurements; 
$A_{sl}^b = -(9.57 \pm 2.51 \pm 1.46) \cdot 10^{-3}~$\cite{Abazov:2010hv} and
$\phi_s^\Delta \in [-0.2,-2.8]~$\cite{combi}
($\phi_s^\Delta \in [0,-1]~$\cite{CDF} at $68\%$ C.F.).
As we see from Fig.~\ref{dphis-Asl}, the theoretical
value for $\phi_s^\Delta$ is consistent with the observation, but
that of $A_{sl}^b$ is still at least a factor of 5 smaller than the D0 measurement
of the dimuon asymmetry \cite{Abazov:2010hv}.

\section{Conclusion}
In this paper we have considered the Higgs sector of a generic supersymmetric extension of the SM, while assuming that there are
more than one pair of $SU(2)_L$ doublet Higgs supermultiplets.
Such a case is realized in models with a low-energy flavor symmetry.
We have calculated  the one-loop effects of the extra Higgs multiplets
to the soft mass insertions. 
We have found  that under a certain circumstance 
the loop effects can give rise to large contributions to the soft 
mass insertions, which means that they can generate large FCNCs and CP violations.
We have applied  a supersymmetric extension of the SM
 based on the discrete $Q_6$ family symmetry.
 Due to the flavor symmetry, the flavor-non-diagonal loop contributions are finite in 
 this model.
We have calculated the supersymmetric contribution
to the non-diagonal matrix element $M_{12}$ 
of the neutral meson systems.
 In particular, we have calculated  the extra phases $\phi_{d,s}^\Delta$
 and the flavor-specific CP-asymmetries $a_{sl}^{d,s}$ in the $B^0$ mixing
 and that the value of $\beta_s$ of the model is consistent 
 with the recent CDF measurement \cite{CDF}.  As for the same sign dimuon
 asymmetry $A_{sl}^b$ we obtain values which are
 one order of magnitude larger than the SM model value.
 Nevertheless, they are at least a factor of 5 smaller than the D0 measurement of the dimuon asymmetry
 \cite{Abazov:2010hv}.

\vspace*{5mm}
J.~K. is partially supported by a Grant-in-Aid for Scientific
Research (C) from Japan Society for Promotion of Science (No.22540271).

\end{document}